\documentclass[11pt,notoc]{JHEP3}
\usepackage{epsfig}

\def\Tr{\mbox{Tr}\;}
\def\MSbar{\overline{\rm MS}}
\def\VEV#1{\left\langle #1\right\rangle}
\newcommand{\ba}         {\begin{eqnarray}}
\newcommand{\ea}         {\end  {eqnarray}}
\newcommand{\be}{\begin{equation}}
\newcommand{\ee}{\end{equation}}
\newcommand{\nn}{\nonumber\\}
\def\({\left(}
\def\){\right)}
\newcommand{\eq}[1]{Eq.~(\ref{#1})}
\newcommand{\fig}[1]{Figure~\ref{#1}}
\newcommand{\tab}[1]{Table~\ref{#1}}
\newcommand{\sect}[1]{Section~\ref{#1}}
\newcommand{\reference}[1]{Ref.~\cite{#1}}

\title{3-d Lattice QCD Free Energy to Four Loops}

\author{F.\ Di Renzo, A.\ Mantovi, V.\ Miccio\\ Dipartimento di Fisica,
  Universit\`a di Parma \\ {\em and}\\ INFN, Gruppo Collegato di Parma, Parma, Italy}
\author{Y.\ Schr\"{o}der\\Center for Theoretical Physics, MIT, Cambridge, MA, USA}

\abstract{
We compute the expansion of the 3-d Lattice QCD free energy to four--loop order by means of Numerical Stochastic Perturbation Theory. 
The first and second order are already known and are correctly reproduced. 
The third- and fourth--order coefficients are new results. 
The known logarithmic divergence in the fourth order is correctly identified. 
We comment on the relevance of our computation in the context of dimensionally reduced finite temperature QCD.
}

\preprint{MIT-CTP-3475\\UPRF-2004-03}


\begin{document}

\section{Introduction}

The QCD free energy density (or the pressure of the quark-gluon plasma) is a good observable to study the deconfinement phase transition \cite{QCDphaseTr}. 
The goal is to study the transition between the realm of low--temperature
hadronic matter, where confinement is the main physical phenomenon, and the
quark-gluon plasma phase that is realized at high temperatures, which in turn
is governed by asymptotic freedom.
In the latter phase the pressure is given by the Stefan--Boltzmann limit of an ideal gas of non--interacting particles, $p \propto T^4$. 
Ideally, one would like to undertake lattice simulations across the phase transition up temperatures at which the pressure exhibits a purely perturbative behavior. 
In practice, however, the convergence properties of the perturbative expansion are poor at temperatures which are not asymptotically large \cite{QCDserie}, while on the other hand computational resources limit the highest temperatures at which lattice simulations can be performed (a fair limit is some $4\div5$ times the transition temperature $T_c \sim 200 MeV$). 

Dimensional reduction \cite{DimRed} is a strategy to fill the gap one is
facing, and in fact has been applied to the problem in question \cite{York}. 
The setup is as follows.
One starts with the full theory ($4d$ QCD) and as a first step matches this to a $3d \, SU(3)$ gauge theory coupled to a Higgs field in the adjoint representation. 
This theory can then be matched to $3d$ pure gauge $SU(3)$, which captures the
ultrasoft degrees of freedom.  
Both these reductions have been successfully performed in a continuum (perturbative) scheme, \emph{i.e.} $\MSbar$. 
$3d$ pure gauge $SU(3)$ then has to be treated non-perturbatively, the only
practical method being lattice measurements.

In order to incorporate these lattice measurements into the reduction setup,
it is essential to know the relation between the two regularization schemes.
This is the point where Lattice Perturbation Theory (LPT) comes into play.
Due to the superrenormalizable nature of the $3d$ theory, {\em all}
divergences can be computed perturbatively. This allows a clean matching
of the schemes in the continuum.
Computing at high orders in LPT is not a simple task (in the present case we
need $g^8$ order -- note that this means four loops for the free energy,
  but three loops for the plaquette), and that is why we make use of
Numerical Stochastic Perturbation Theory (NSPT) \cite{NSPT}.

We recall the definition of the free energy density $f$ 
\be
	Z = \int DU \, e^{-S_W[U]}\; = e^{ - \frac{V}{T} f} \;,
\ee
where the (Wilson) pure gauge action reads 
\be
	S_W = \beta_0 \sum_P (1 - \Pi_P) \;,
\ee
with $\beta_0=2N_c/(a^{4-d} g_0^2)$ denoting the the standard dimensionless (bare) lattice inverse coupling in $d$ dimensions, while $\Pi_P$ is the basic plaquette
\be
	\Pi_P = \frac{1}{N_c} \Re (\Tr U_P) \;,
\ee
which is to be computed at any point on any independent plane according to 
\be
	U_P = U_{\mu\nu}(n) = U_\mu(n) U_\nu(n+\mu) U^{\dagger}_\mu(n+\nu) U^{\dagger}_\nu(n) \;.
\ee
To compute the free energy one can now revert to the computation of the plaquette
\ba
	\VEV{1 - \Pi_P} & = & Z^{-1} \int DU \, e^{-S_W[U]}\; (1 - \Pi_P)	\\ \nonumber
	& = & - \frac{2\, a^d}{d(d-1)V} \frac{\partial}{\partial \beta_0} \ln Z = \frac{2\, a^d}{d(d-1)} \frac{\partial}{\partial \beta_0} (\frac{1}{T} f) \;.
\ea
Hence, given a weak--coupling expansion of the plaquette 
\be
	\VEV{1 - \Pi_P} = \frac{c_1(N_c,d)}{\beta_0} + \frac{c_2(N_c,d)}{\beta_0^2} + \frac{c_3(N_c,d)}{\beta_0^3} + \frac{c_4(N_c,d)}{\beta_0^4} + \ldots
\label{PLexpns}
\ee
it follows that
\be
	\frac{2\, a^d}{d(d-1)}(\frac{1}{T} f) = c_0(N_c,d) + c_1(N_c,d) \ln \beta_0 - \frac{c_2(N_c,d)}{\beta_0} - \frac{c_3(N_c,d)}{2 \beta_0^2} - \frac{c_4(N_c,d)}{3 \beta_0^3} - \ldots \;.
\ee
We now specialize to QCD ($N_c=3$) in $d=3$ dimensions, where $g_0^2 \sim
a^{-1}$ and hence $\beta_0 = 6/(a g_0^2)$. The previous formula reads (from
here on, $c_i\equiv c_i(N_c=3,d=3)$)
\be
	\frac{2}{6}\, \frac{1}{T} f = a^{-3} \( c_0 + c_1 \ln \beta_0 \) - a^{-2} \frac{c_2}{6} g_0^2 - a^{-1} \frac{c_3}{72} g_0^4 - \frac{\tilde{c}_4}{648} g_0^6 + \emph{O}(a) \;.
\label{PLtld}
\ee
In order to control the matching to continuum one then needs the first four coefficients in the expansion of the basic plaquette. Note however that it is already known from a computation in the continuum that at four loop level there is a logarithmic infrared (IR) divergence \cite{York4L}. 
One of the aims of our computation is to recover the scheme--independent coefficient of this logarithm, while fixing the lattice constant which is left over once an IR regulator has been chosen. 
In \eq{PLtld} we have put a tilde on $c_4$ to denote that the IR divergence has to be isolated and subtracted in a convenient scheme. Later on, the lattice size $L$ will act as the IR regulator. 
Going back to the lattice coefficients themselves, the first and the second ones are already known \cite{HK}. We will give them in \sect{sec:Results}. 
The third and the fourth ones are the goal of the present work, a task which one can manage within our computational scheme\footnote{In $4d$ the expansion of the plaquette is known via NSPT up to a much higher order. 
It is interesting to compare the two different situations. 
In $3d$ there is the additional subtlety of the IR divergence. 
On the other hand, the dimensionful nature of the coupling in $3d$ makes it possible to single out the different divergent contributions in Perturbation Theory. 
The situation is much more involved in $4d$ (see \cite{4d10L}).}.


\section{Computational setup} \label{sec:Setup}

Our computational tool is NSPT. Computing to $\beta_0^{-4}$ order requires to expand the field up to $\beta^{-4}$ order \cite{NSPT}, that is 
\be
	U_\mu(n) = 1 + \sum_{i=1}^{8} \, \beta_0^{-\frac{i}{2}} \; U_\mu^{(i)}(n) \;.
\label{Uexpns}
\ee
We write the expansion in terms of the $U_\mu$ field. 
One could also express everything in terms of the Lie algebra field $A_\mu(n) = \sum_{i=1}^{8} \beta_0^{-\frac{i}{2}} A_\mu^{(i)}(n)$, the relation being $U_\mu(n) = \exp(A_\mu(n))$. 
Whichever one uses, one should keep in mind that perturbation theory amounts in any case to decompactify the formulation of lattice gauge theory. 
The expansion in terms of the $U_\mu^{(i)}(n)$ is easier to manage from the point of view of computer data organization. 
\eq{Uexpns} is the expansion to be inserted in the Langevin equation
\be
	\partial_tU_\eta=[-i\nabla S[U_\eta]-i\eta]U_\eta \;,
\ee
$\eta$ being a Gaussian noise. 
The equation has to be integrated in a convenient (time) discretization scheme. 
Our choice is the Euler scheme as it was proposed in \cite{Batr}. 
This amounts to introducing a time step $\epsilon$. 
As usual, the solution is recovered by working in a region where the time step corrections are linear (Euler scheme is a first order scheme) and extra\-polating to $\epsilon \rightarrow 0$. 
We computed the expansion in \eq{PLexpns} for $\VEV{1 - \Pi_P}$ on different lattice sizes ranging from $L=5$ to $L=16$ (up to three loops we also performed the computations on a $L=18$ lattice) 
\be
	\VEV{1 - \Pi_P}^{(L)} = \frac{c_1^{(L)}}{\beta_0} + \frac{c_2^{(L)}}{\beta_0^2} + \frac{c_3^{(L)}}{\beta_0^3} + \frac{c_4^{(L)}}{\beta_0^4} + \ldots \;.
\label{PLexpnsL}
\ee
We then extrapolated the infinite lattice size results according to 
\ba \label{FV}
	c_1^{(L)} & = & c_1 + \frac{d_1}{L^3}  \nn
	c_i^{(L)} & = & c_i + \sum_{j=j_i}^{J_i} \sum_{k=0}^{i-1} \, d_i^{(j,k)} \,\, \frac{\ln^k L}{L^j} \qquad \qquad \qquad (i=2,3) \\ 
	c_4^{(L)} & = & c_4^{(ln)} \ln L^3 + c_4 + \sum_{j=j_4}^{J_4} \sum_{k=0}^{3} \, d_4^{(j,k)} \,\, \frac{\ln^k L}{L^j} \;. \nonumber
\ea
These asymptotic forms are basically dictated by Symanzik's analysis
\cite{LW}. In particular they include the contribution from subleading
logarithms (they are suppressed by inverse powers of $L$). For each $i$ the
index $k$ runs up to $i-1$, which equals the number of loops (remember that in
terms of the plaquette we are computing up to three loops, \emph{i.e.} $i=1$
is the tree level). The index $j$ counts the subleading contributions coming
from inverse powers of $L$. As it appears from the last line of the previous
formula, the finite volume also acts as the IR regulator needed at order
$\beta_0^{-4}$ (this is instead a leading logarithm). Some comments are in
order at this point. The final errors on infinite volume results are dominated
by this extrapolation process. 

Trying to assess the effect of the subleading logarithms, it turns out that both the range of our data and our statistical errors do not allow to distinguish between a logarithm and a constant. Hence we will only give (effective) extrapolations based on pure power-like fits. The spread of the results comes from the indetermination on the (inverse) powers to be included in the fit. This is not surprising, since in NSPT there is no control on what in the language of Feynman diagrams would be contributions coming from different diagrams (\emph{i.e.} sums). We then try different choices of the powers and then compare the corresponding $\chi^2$'s. This process does not select a definite set of powers: the better choices (see \fig{fig:ResFig2}) turn out to be comparable with respect to the resulting $\chi^2$. The quoted values for the $c_i$ ($i>1$) together with the associated errors embrace the range of the outcomes. 

Most of the computer simulations have been performed on a PC cluster the Parma group installed one year ago. 
This is made of ten bi-processor Athlon MP2200. 
A programming environment for NSPT for Lattice Gauge Theory was set up in C++. 
This was in part inspired by the TAO codes we use on the APE machines and for a large fraction based on the use of (C++ specific) {\it classes} and {\it methods} to handle lattice and algebraic structures. 
Needless to say, this part of the work will be useful in other applications of NSPT. 
The results we report come out of 6 months of runs on the above mentioned cluster. 
Some more statistics came from another PC cluster more recently installed in Parma. 
The latter is a \emph{blade} system based on 14 Intel Xeon 2.0 GHz.


\section{Results} \label{sec:Results}

\TABULAR[t]
{c|c|c|c|c c}
{\hline
$L$   & $c^{(L)}_1$ & $c^{(L)}_2$ & $c^{(L)}_3$ & $c^{(L)}_4$ & $c^{(L)}_4 - c_4^{(ln)} \ln L^3$ \\
\hline
$5$              & $2.6455(13)$ & $1.8682(45)$ & $5.990(26)$ & $25.99(18)$ & $21.28(18)$ \\ 
$6$              & $2.6536(8)$ & $1.8968(31)$ & $6.200(19)$ & $27.66(14)$  & $22.41(14)$ \\
$7$              & $2.6580(8)$ & $1.9095(30)$ & $6.307(21)$ & $28.68(15)$  & $22.98(15)$ \\
$8$              & $2.6615(6)$ & $1.9226(23)$ & $6.408(16)$ & $29.66(14)$  & $23.57(14)$ \\
$9$              & $2.6630(6)$ & $1.9288(22)$ & $6.484(18)$ & $30.44(16)$  & $24.00(16)$ \\
$10$             & $2.6638(4)$ & $1.9340(15)$ & $6.519(13)$ & $30.91(13)$  & $24.16(13)$ \\
$11$             & $2.6645(4)$ & $1.9381(14)$ & $6.574(11)$ & $31.53(14)$  & $24.51(14)$ \\
$12$             & $2.6650(3)$ & $1.9413(12)$ & $6.591(11)$ & $31.67(15)$  & $24.39(15)$ \\
$13$             & $2.6653(3)$ & $1.9423(12)$ & $6.621(11)$ & $32.27(18)$  & $24.76(18)$ \\
$14$             & $2.6656(3)$ & $1.9436(12)$ & $6.288(11)$ & $32.37(16)$  & $24.64(16)$ \\
$15$             & $2.6662(2)$ & $1.9455(10)$ & $6.652(10)$ & $32.84(19)$  & $24.91(19)$ \\
$16$             & $2.6657(2)$ & $1.9442(8)$ & $6.658(9)$ & $33.28(19)$  & $25.16(19)$ \\
$18$             & $2.6663(2)$ & $1.9489(7)$ & $6.715(8)$ & $-$  & $-$ \\
$$               & $$ & $$ & $$ & $$  & $$ \\
$\infty$      & $2.6666(1)$ & $1.955(2)$ & $6.90^{(2)}_{(12)}$ & $ $  & $25.8(4)$ \\
\hline}
{\label{tab:ResTable}The coefficients $c_i^{(L)}$ at the various lattice sizes and their infinite volume extrapolations $c_i$. For the last order we report both $c_4^{(L)}$ and $c_4^{(L)} - c_4^{(ln)}\ln L^3$; the latter is the quantity to be extrapolated. For the error on $c_3$ see text and \fig{fig:ResFig2}.}

In \tab{tab:ResTable} we present the results we obtained for the coefficients
$c_i^{(L)}$ at various values of $L$. As in \eq{PLexpns} and \eq{PLexpnsL}, we only give the coefficients of order $\beta_0^{-n}$: the coefficients odd in $g_0$ (\emph{i.e.} of order $\beta_0^{-(2n+1)/2}$) were verified to be zero within errors. In the last line one can read the values extrapolated to infinite volume. For the fourth order we present both the bare coefficients $c_4^{(L)}$ and the subtracted ones $c_4^{(L)} - c_4^{(ln)}\ln L^3$; the latter is the finite quantity one is interested in at $L=\infty$. 

\EPSFIGURE[t]
{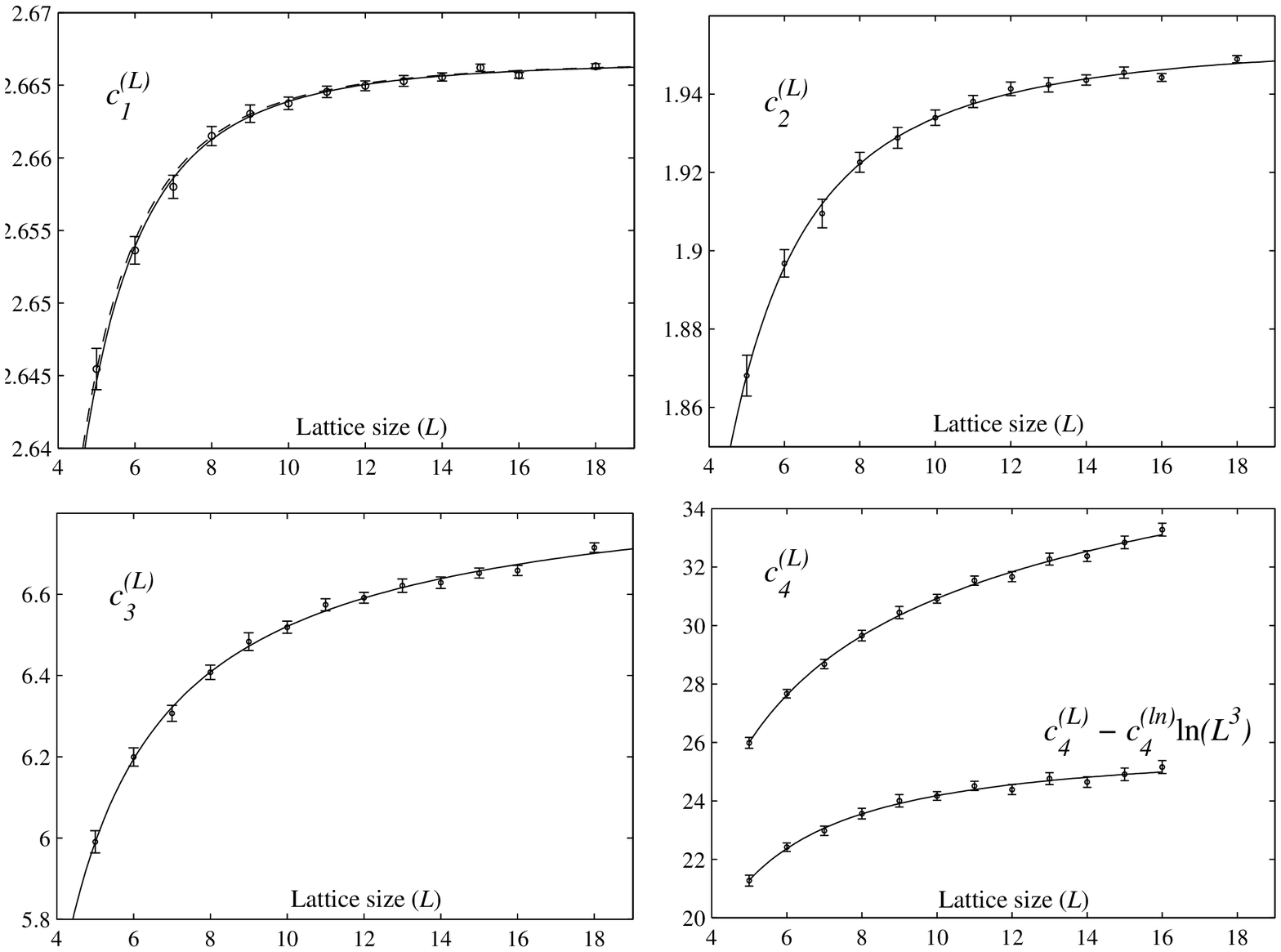,scale=0.73}
{\label{fig:ResFig} The coefficients $c_i^{(L)}$ together with
	the interpolating finite size corrections (solid lines). For first
	order we also plot the (known) analytic finite volume corrections
	(dashed line). For the last order we plot both $c_4^{(L)}$ and
	$c_4^{(L)} - c_4^{(ln)}\ln L^3$.}

In \fig{fig:ResFig} we plot the values of the coefficients at the various lattice sizes together with the interpolating finite size corrections. Again, for the fourth order we plot both the bare coefficients $c_4^{(L)}$ and the subtracted ones $c_4^{(L)} - c_4^{(ln)}\ln L^3$. An obvious benchmark for our computations is the first order, whose value is $c_1^{(L)} = 8/3*(1-1/L^3)$. 
In this (trivial) case one knows the result both at finite and at infinite volume. 
That is why in \fig{fig:ResFig} we also plot the known finite size corrections for $c_1$. 

Another benchmark is the second coefficient, which is also found in agreement
with the diagrammatic studies in \cite{HK}. From \reference{Kguess} one reads
$c_2 = 1.9486$, however without an error estimate. 

As it was already pointed
out, for orders higher than the trivial one, the quoted errors of the infinite
volume--extrapolated values are dominated by the form of the fitting
polynomials in \eq{FV}. Still, the final errors are acceptable. Note the asymmetric error for $c_3$. The more conservative lower bound takes into account a choice for the subleading powers of $L$ which results in a worse $\chi^2$, see \fig{fig:ResFig2}. It is interesting to compare the result $c_3$ (our first original result) with the one conjectured in \cite{Kguess} from the hypothesis of the dominance of a given contribution: $c_3 \approx 7.02$. This conjecture turned out to be not too far from the result. 

\DOUBLEFIGURE[t]
{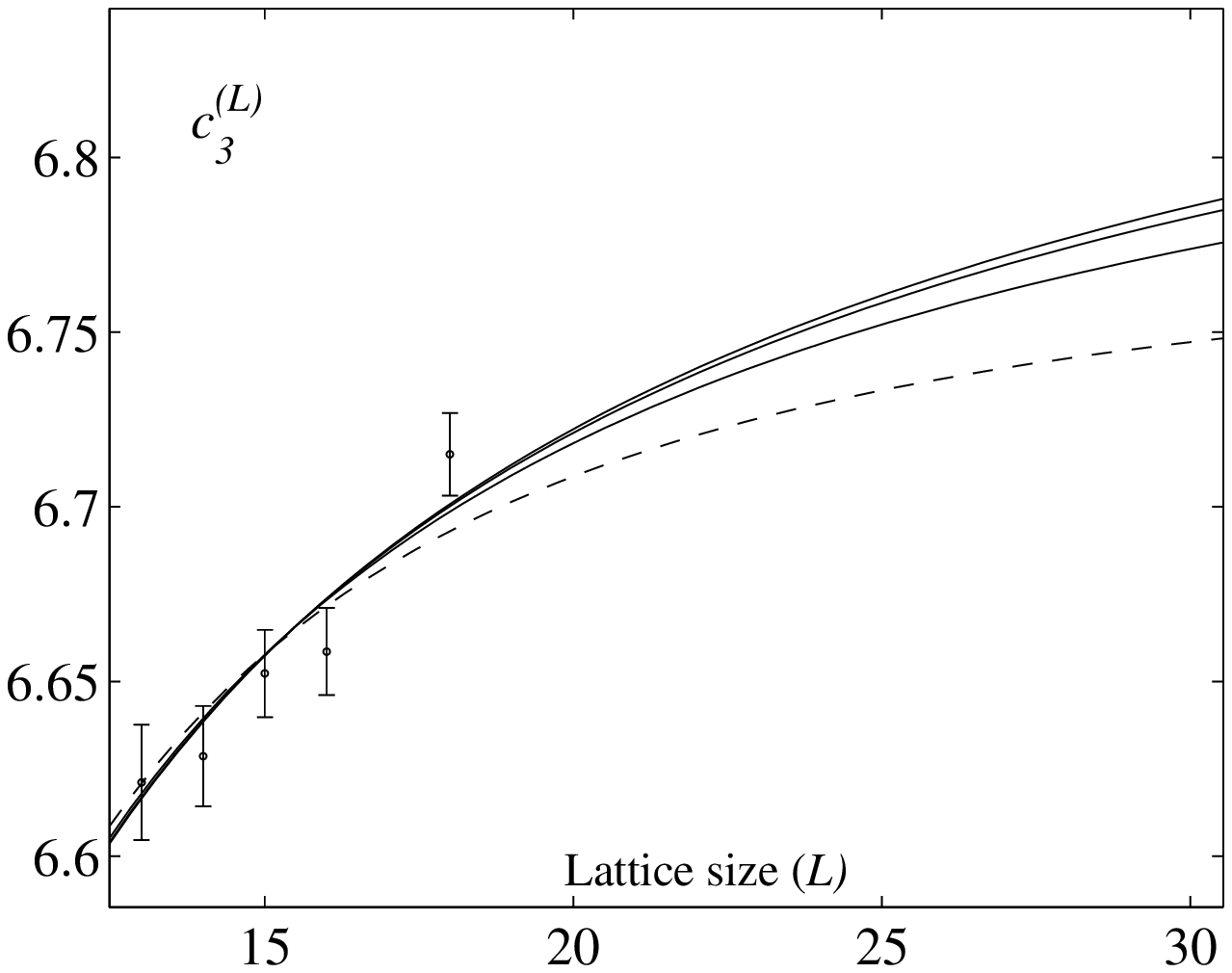,scale=0.49}
{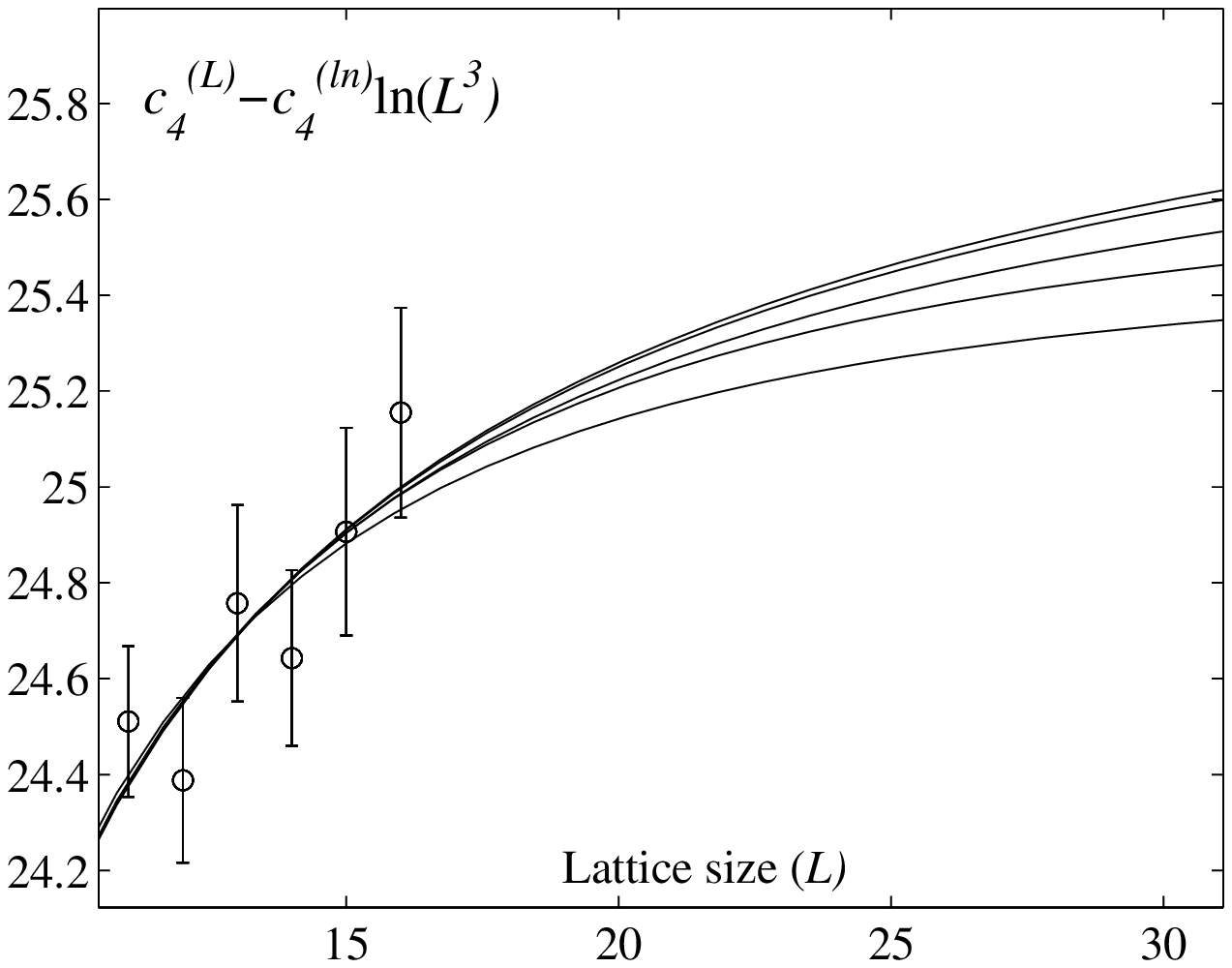,scale=0.5}
{\label{fig:ResFig2} Different fits for $c_3^{(L)}$. The upper curves takes $1/L$ as the leading power, while the lower (dashed) one takes $1/L^2$. Note that the latter results in a higher $\chi^2$ ($0.7$ vs $0.55$).}
{\label{fig:ResFig3} Different fits for $c_4^{(L)} - c_4^{(ln)}\ln L^3$. In these fits the value for $c_4^{(ln)}$ is fixed to be the analytically known one. Within the range of the data the fits almost coincide.}

Let us now discuss the IR divergence at order $\beta_0^{-4}$. As is well known,  it is difficult to fit a logarithm. Still, we obtain enough evidence for it. By this we mean the following. One can take different approaches to the fit of the last line of \eq{FV}. One possibility is to include no logarithmic correction at all. A second one is to include a logarithmic correction whose prefactor $c_4^{(ln)}$ is a fitting parameter. A third possibility is to include a logarithmic correction whose prefactor $c_4^{(ln)}$ equals the result which has already been obtained in the continuum computation of \cite{York4L}: $c_4^{(ln)} = 81\,(688-157\pi^2/4)/(4\pi)^4  = 0.9765$. By varying the choice of the inverse powers included in the fit, the first case (no log) yields values of $\chi^2$ which are systematically worse than in the other two cases: as expected, the data appear to prefer the inclusion of the logarithm. By fitting both $c_4$ and $c_4^{(ln)}$ we obtained $c_4=24.5(2.0)$ and $c_4^{(ln)}=1.1(2)$, a result fully consistent with \cite{York4L}. In the third case (see \fig{fig:ResFig3}) we obtain $c_4=25.8(4)$, getting a smaller error like expected. For the meaning of the quoted errors, see the discussion in \sect{sec:Setup}.


\section{Conclusions and perspectives}

We computed the first four coefficients in the expansion \eq{PLtld} of the plaquette in $3d$ pure gauge $SU(3)$ theory, from which one can trivially obtain the expansion of the free energy at four loops. 
For the first two coefficients the already known results have been correctly reproduced. 
The third coefficient (the first original result of this paper) is connected with the mildest power divergence to be subtracted from simulations data for a lattice determination of the free energy. 

The fourth coefficient was known to be logarithmically divergent, a result that was reproduced. 
Let us further comment on this point. 
A priori one can use any IR regulator in order to extract a finite part for the four loop contribution: the finite volume (the one we used in the present work), a mass (a very popular IR cutoff in the continuum) or the coupling itself (since it is dimensionful in $3d$). 
Notice that the latter is in a sense the natural choice for computer simulations. 
Obviously each choice defines a scheme of its own. 
While the coefficient of the logarithm is universal, there are of course specific constants relating the different schemes. 
As already pointed out, the coupling itself is the most natural regulator for computer simulations, even if there is no simple way to take it as the cutoff in perturbation theory. 
Ultimately, we are interested in the matching between the lattice and a
continuum perturbative scheme (to be definite, $\MSbar$).
The idea is to take the same IR regulator both in continuum and in lattice
perturbation theory, which most naturally would be a common mass for all
tree-level propagators. 
Since the same mismatch will be present in both computations with respect to the data coming from computer simulations, that mismatch will cancel in the matching. 
Employing massive propagators in NSPT will therefore be the natural extension
of the approach presented here.


\acknowledgments

We thank C. Torrero for having collaborated with us in data analysis. The Parma group acknowledges support from MIUR under contract 2001021158 and from I.N.F.N. under {\sl i.s. MI11}. Y.S. acknowledges support from the DOE, under Cooperative Agreement no.~DF-FC02-94ER40818. F.D.R. and Y.S. also acknowledge support from the {\sl Bruno Rossi INFN-MIT exchange program}.


\end{document}